# A Brief Comparison of Two Enterprise-Class RDBMSs

Andrew Figueroa
Department of Computer Science
Western Connecticut State University
181 White St. Danbury, CT 06810
figueroa039@connect.wcsu.edu

Steven Rollo
Department of Computer Science
Western Connecticut State University
181 White St. Danbury, CT 06810
rollo003@connect.wcsu.edu

Sean Murthy
Department of Computer Science
Western Connecticut State University
181 White St. Danbury, CT 06810
murthys@wcsu.edu

## ABSTRACT

This paper is an extended version of a report from a student-developed study to compare Microsoft® SQL Server® and PostgreSQL, two widely-used enterprise-class relational database management systems (RDBMSs). The study followed an introductory undergraduate course in relational systems and was designed to help gain practical understanding of specific DBMSs. During this study, we implemented three non-trivial schemas in each system, identified 26 common database design, development, and administration activities while implementing the schemas, and compared the support each system offers to carry out the identified activities. Where relevant, we also compared each system against the SQL standard.

In this report, we present a summary of the similarities and differences we found between the two systems, and we provide a quantitative measure ranking both systems' implementations of the 26 activities. We also briefly discuss the "technical suitability" of PostgreSQL to enterprise applications. Although this report is not comprehensive and is too general to comment on the suitability of either system to a specific enterprise application, it can nevertheless provide an initial set of considerations and criteria to choose a system for most enterprise applications.

## Categories and Subject Descriptors
H.2.4 [**Database Management**]: Systems – *Relational databases.*

## 1. INTRODUCTION
For an undergraduate research project, we sought to gain a better practical understanding of relational database management systems (RDBMSs) used to build enterprise applications. Specifically, we studied Microsoft® SQL Server® 2016 Express Version 13.0.4001 (MSSQL) and PostgreSQL Version 9.6.3 (Postgres), and their respective SQL implementations Transact-SQL (T-SQL) and PL/pgSQL (pgSQL).

Our study proceeded in three broad steps: implementing schemas and queries for three non-trivial applications in each selected DBMS, identifying common database-related activities, and comparing the facilities each DBMS offers to carry out the identified activities.

We began the study by porting to both MSSQL and Postgres the following schemas originally designed for Oracle® Database:

**Advert**: This schema describes an advertisement management system for a fictional newspaper.

**Shelter**: This schema describes a system for managing animal care and adoptions at a fictional animal shelter.

**Babysitting**: This schema describes a fictional babysitting co-op where members earn and use credits through babysitting each other's children.

**Classroom**: This scenario represents an environment where students in a course are allowed appropriate access to a DBMS to experiment and complete assignments. Instructors for this course also have access to the student's work inside of the DBMS. This scenario originates from our work on the open source ClassDB project [3].

As we ported the schemas and queries, we observed 26 distinct common database activities. We categorized each distinct activity as either a design, development, or administration activity. We considered as *design* any feature that influences how a schema is created at the logical or physical levels (for example, data types and indexes), whereas we classified as *development* any direct management and manipulation of data ("DML queries"). We labeled activities involving the operation and maintenance of the DBMS as *administration*. While many activities examined in this study could be placed in more than one category, we placed activities into the category that best represented the intent behind the activity. For example, we considered writing triggers an administration activity instead of a development activity because triggers are often used to perform administrative tasks.

With a list of database activities and a classification in hand, we then studied the details of how MSSQL and Postgres support the activities and compared the approaches the two systems take. Where relevant, we compared each system against the SQL:1999 standard ("the SQL standard") [7]. We also assigned a quantitative score for each activity to both DBMSs based on the level of support and adherence to the SQL standard.

At a high-level, we conclude from this study the following: Both MSSQL and Postgres include competitive features to support common database activities involved in building enterprise systems, with MSSQL taking a more "implementer friendly" approach. Postgres is quite suitable (technically speaking) for enterprise applications, but its lack of certain features may mean some activities require additional implementation effort. Lastly, Postgres's SQL implementation is very close to the SQL standard, whereas MSSQL's is not. However, standards-compliant code is often not portable and at times can be quite inefficient.

The rest of this paper is organized as follows: Sections 2, 3, and 4 introduce the design, development, and administration activities, respectively. These sections detail the differences found between the two DBMSs for each activity. Section 5 discusses the differences found between MSSQL and Postgres, and provides a quantitative score for both systems. Section 6 presents a summary of the topics covered in this paper, and discusses the technical suitability of Postgres to enterprise applications. The SQL implementations of the schemas for both MSSQL and Postgres, as well as the examples in this paper, are all available in a public Git repository [2].

| SQL:1999 | Microsoft SQL Server | PostgreSQL |
|---|---|---|
| `INTEGER` | `INTEGER` (4 bytes) | `INTEGER` (4 bytes) |
| `SMALLINT` | `SMALLINT` (2 bytes) | `SMALLINT` (2 bytes) |
| `BIGINT` | `BIGINT` (8 bytes); `TINYINT` (1 byte; unsigned) | `BIGINT` (8 bytes) |
| `NUMERIC` | `NUMERIC` | `NUMERIC` |
| `DECIMAL` | `DECIMAL` (up to 38 digits total before and after decimal pt.) | `DECIMAL` (synonymous with `NUMERIC`) (131072 digits before decimal pt.; 16383 after) |
| `REAL` | `REAL` (same as `FLOAT(24)`) | `REAL` (IEEE single precision) |
| `DOUBLE PRECISION` | `FLOAT(n)` | `DOUBLE PRECISION` (IEEE double) |
| `FLOAT` | (1-24: 7-digit precision; 25-53: 15-digit precision) | `FLOAT(n)` (1-24: single; 25-53: double) |
| `CHAR` | `CHAR` | `CHAR` |
| `VARCHAR` | `VARCHAR` | `VARCHAR` |
| `CLOB` | `VARCHAR(max)` (approximately 2GB maximum) | `TEXT` (approximately 1GB maximum) |
| `DATE` (year 0001-9999) | `DATE` (year 0001-9999; 3 bytes) | `DATE` (year 4713 BCE to 5874897 CE; 4 bytes) |
| `TIME` (6 fractional secs) | `TIME` (7 fractional secs; 3-5 bytes) | `TIME` (6 fs; 8 or 12 bytes; includes time zone) |
| `TIMESTAMP` (year 0001-9999; 6 fs) | `DATETIME` (year 1753-9999; 3 fs; 8 bytes) `DATETIME2` (year 0001-9999; 7 fs; 6-8 bytes) `SMALLDATETIME` (year 1900-2079; 0 fs; 4 bytes) `DATETIMEOFFSET` (year 0001-9999; 7 fs; 8-10 bytes; time zone) | `TIMESTAMP` (year 4713 BCE to 294276 CE; 6 fs; 8 bytes; time zone) |

**Table 1. Common SQL:1999 data types and their implementation. All numeric types except `TINYINT` are signed. SQL:1999 does not define storage size for any type and defines value ranges only for date and time types.**

## 2. DESIGN ACTIVITIES

In this section, we compare how MSSQL and Postgres support various database design activities.

### 2.1 Field Design

We studied several aspects of field design, chief among them being data types related to three common kinds of data: numeric, character, and date-time. Table 1 summarizes the relevant data types defined in the SQL standard and their equivalents in MSSQL and Postgres.

**Numeric fields:** The SQL standard describes two categories of numeric data: *exact numerics* and *approximate numerics*. An exact numeric is a number defined with a specific maximum number of digits. The types `INTEGER`, `SMALLINT`, and `BIGINT` represent integral numbers. The types `NUMERIC` and `DECIMAL` represent numbers with fractional parts and permit the designer to choose both *precision* (total number of digits) and *scale* (number of fractional digits).

The standard permits the precision and accuracy of numeric types to be implementation-defined, but requires that the precision of `SMALLINT` be less than or equal to that of `INTEGER` and precision of `INTEGER` be less than or equal to that of `BIGINT`. It also states that `NUMERIC` must always be represented with the exact precision requested, while `DECIMAL` can be managed with a precision greater than what was specified. Another restriction is that `DOUBLE PRECISION` should be of a greater precision than that of the `REAL` type. However, the standard does not specify how much greater the precision should be.

Approximate numerics include `REAL`, `DOUBLE PRECISION`, and `FLOAT` [8§2.4] and are generally implemented as floating-point numbers. The available range and accuracy (point at which preciseness is lost due to limitations of their representation) of `REAL` and `DOUBLE PRECISION` are implementation defined and fixed, whereas the `FLOAT` type is designed to allow designers to specify the desired precision.

Both MSSQL and Postgres fully support the SQL standard for exact numeric types, but MSSQL also adds a `TINYINT` type for integral values ranging from 0 to 255. Both systems treat the `NUMERIC` and `DECIMAL` data types as synonyms, but they afford different maximum precision [19][27][54].

For approximate numerics, MSSQL does not define a `DOUBLE PRECISION` data type, but does provide `REAL` and `FLOAT` types. This adaptation does not change the precision or range available because the `FLOAT` type provided by MSSQL has a maximum range identical to that of the `DOUBLE PRECISION` type Postgres provides (when using the IEEE 754 standard in Postgres) [22].

Postgres implements all three approximate numeric types that SQL:1999 defines. In most cases, Postgres represents both the `REAL` and the `DOUBLE PRECISION` types as IEEE Standard 754 Binary Floating-Point numbers [6], representing the single-precision and double-precision parts of the standard, respectively. This is the expected behavior so long as "the underlying processor, operating system, and compiler support it" [54]. Postgres maps `FLOAT` to the other two approximate numeric types, depending on the precision specified. `FLOAT(1)` to `FLOAT(24)` results in the `REAL` type and `FLOAT(25)` to `FLOAT(53)`, or `FLOAT` without a precision specified results in Postgres selecting the `DOUBLE PRECISION` type.

**Character fields:** SQL:1999 defines three different character types: `CHAR`, `VARCHAR`, and `CLOB`. `CHAR` values are fixed length (right-padded with spaces), whereas `VARCHAR` fields have a variable length, with a user specified maximum length, after which a string is truncated. The `CLOB` type (character large object) is meant to store large strings of characters without a predetermined maximum length.

Both MSSQL and Postgres implement the `CHAR` and `VARCHAR` types as defined by the standard. Both also effectively implement the `CLOB` type, but use a different name. MSSQL uses the `VARCHAR(max)` data type and Postgres uses the `TEXT` data type.

**Date and time fields:** The SQL standard defines three basic date and time types: `DATE`, `TIME`, and `TIMESTAMP` [8§2.4]. MSSQL closely follows the SQL recommendation for the `DATE` and `TIME` types with years ranging from 0001 to 9999 C.E. for `DATE` and 7 fractional seconds for `TIME`. However, MSSQL's implementation uses `DATETIME` as a counterpart to `TIMESTAMP`. This is an important distinction because MSSQL does have a data type named `TIMESTAMP`, but is used for an entirely unrelated purpose. This can lead to issues when naively porting a schema that uses `TIMESTAMP` to MSSQL [39]. Additionally, MSSQL's `DATETIME` is not completely equivalent to the standard's `TIMESTAMP`, as it can only store years from 1753-9999, and only allows up to 3 digits for fractional seconds (compared to 6 in both the SQL standard and Postgres).

In addition to these three types, MSSQL also has other types available for storing date and time. These are `SMALLDATETIME` (years 1900-2079, 1 minute precision, 4 bytes), `DATETIME2` (years 0001-9999, 6 fractional seconds, 6-8 bytes) and `DATETIMEOFFSET`, which is identical in precision and range to `DATETIME2`, but allows a UTC time offset (time zone) [18]. Although it breaks backwards compatibility with versions prior to SQL Server 2008, it is recommended to use `DATETIME2` over `DATETIME` whenever possible, as it has an expanded range and precision, and may use less storage space, depending on the precision of the value stored.

In contrast, Postgres fully implements all three date-time types the standard defines, along with `TIME WITH TIME ZONE` and `TIMESTAMP WITH TIME ZONE` derivatives. One deviation from the standard is the range of years able to be stored in the `DATE` and `TIMESTAMP` types: The standard states the year value is to be exactly 4 digits in length, with years ranging from 0001 to 9999. However, Postgres supports years ranging from 4713 BCE to 5874897 CE (up to 294276 CE with `TIMESTAMP`) [55].

## 2.2 Computed Columns

Computed columns are not specified in the standard, but serve as a convenient design abstraction because they provide a way to represent the results of an expression as a column, rather than repeating the expression in every related query.

MSSQL natively supports computed columns. Postgres does not provide this feature, requiring the use of a view to add the computed column. However, as described in section 2.4, views impose constraints on inserts and edits. An alternative is to add "read only" columns and use triggers to update column values. However, making columns read only and adding triggers requires much additional administration and design effort. Figure 1 shows an example implementation of computed columns in MSSQL and Postgres (via a view).

In MSSQL, computed columns can consist of any expression that does not involve a query. The expressions that Postgres allows in a computed column depend on how the computed column is being implemented. When using a view, any expression that would normally be permitted as a part of a `SELECT` statement on the underlying table can be used. In this instance, computed columns in both DBMSs are only allowed to reference the columns of the table where it resides, and not those of other tables. Other behavior is also similar between MSSQL's computed columns and Postgres' implementation using a view. This includes characteristics such as the inability to modify the computed columns directly using `INSERT` or `UPDATE` statements, and that they are virtual (not stored) by default [32].

```
CREATE TABLE Dog (
  dog_id INTEGER PRIMARY KEY,
  name VARCHAR(15),
  weight NUMERIC(4,1),
  weightKg AS weight * 0.45359 );
```
**Figure 1a. Implementing a computed column in MSSQL**

```
CREATE TABLE Dog (
  dog_id INTEGER PRIMARY KEY,
  name VARCHAR(15),
  weight NUMERIC(4,1) );

CREATE VIEW Dog_Kg AS (
  SELECT dog_id, name, weight,
    (weight * 0.45359) AS weightKg
  FROM dog );
```
**Figure 1b. Implementing similar, but not identical functionality in Postgres**

## 2.3 Index Design

Despite not being defined in the SQL standard, indexes comprise a significant portion of schema design. They provide necessary optimizations and are frequently used by internal DBMS operations. However, this lack of a specification by the SQL standard does not necessarily mean that their implementations differ greatly. Rather, both DBMSs implement indexes in a relatively similar manner. Apart from some syntactic differences, the creation and behavior of indexes in both DBMSs is similar, meaning that most indexes can be designed without needing much modification when moving from one DBMS to the other.

In both DBMSs, changing the type of index affects the optimal data structures and algorithms associated with each index. One difference that could be considered is that MSSQL stores most "regular" indexes using a B-Tree storage method. This method can only be modified by changing the type of index [25]. Postgres instead allows the internal storage method to be manually set for each index. The storage methods for indexes provided by Postgres include B-tree, hash, GiST, SP-GiST, GIN, and BRIN [44]. Additionally, Postgres allows defining a custom index storage method, but with a warning that the process can be rather difficult.

Some differences may arise when needing to create special kinds of indexes. The types we studied include: indexes with included columns, filtered indexes, and indexes on expressions.

**Included Columns:** Indexes with included columns are indexes where in addition to the key attributes, one or more columns are also included and stored with the index. However, these extra columns are not used to organize the index, making them "non-key" columns in relation to the index. By adding included columns, a well-designed index can eliminate the need for reading data pages for a significant portion of the workload, greatly improving performance for queries that benefit from this covering index. In MSSQL this is performed by using the `INCLUDE` keyword (see Figure 2a) [13]. Adding a column in this manner reduces the cost of maintaining the index compared to adding it as a key column for the index. Although Postgres supports indexes with more than one column, it does not support included columns. Instead, additional columns must be added as key columns for the index (Figure 2b) [45]. This reduces the net benefit of indexes with multiple columns, as additional computation is needed to maintain the sort order for the index.

```
CREATE INDEX IX_Dog_ArrivalDate
ON dog (arrival_date)
INCLUDE (name, breed);
```
**Figure 2a. An index with included columns in MSSQL**

```
CREATE INDEX IX_Dog_ArrivalDate
ON dog (arrival_date, name, breed);
```
**Figure 2b. A multi-column index in Postgres**

**Filtered Indexes:** A filtered index includes only selected rows of the table that it is indexing. Including only selected rows can significantly reduce the size and the computational overhead caused by the index, since it does not need to be updated for rows that would not benefit from being indexed. These filtered indexes are especially useful for tables that have many records and would benefit from indexing, but only a small portion are frequently accessed. The rows in the index are selected through the value of a specified conditional statement. Both MSSQL [11] and Postgres [47] use a similar syntax for creating filtered indexes. Figure 3 shows a filtered index being created in both DBMSs.

```
CREATE INDEX IX_Dog_RecentArrival
WHERE arrival_date > '2010-01-01';
```
**Figure 3. A filtered index in both MSSQL and Postgres**

**Indexing on Expressions:** Indexes can also be created on expressions which involve the columns of a table. These are particularly useful for queries that involve sorting or grouping by an expression, especially if the expression in question is computationally expensive and frequently used. MSSQL does not allow indexes on expressions, but it does allow them on computed columns. [26]. In contrast, Postgres directly allows indexes on expressions [46], but as noted earlier, does not support computed columns. See Figures 4a and 4b for an illustration of the differences in implementation.

```
CREATE TABLE Volunteer (
  vol_id INTEGER PRIMARY KEY,
  fullName VARCHAR(50)
  fullNameCaps AS UPPER(fullName) );

CREATE INDEX IX_Volunteer
ON Volunteer(fullNameCaps);
```
**Figure 4a. An index on an expression in MSSQL**

```
CREATE TABLE Volunteer (
  vol_id INTEGER PRIMARY KEY,
  fullName VARCHAR(50) );

CREATE INDEX IX_Volunteer
ON Volunteer(UPPER(fullName));
```
**Figure 4b. An index on an expression in Postgres**

## 2.4 Views

One common requirement when designing a database application is to provide a simplified way to access the results of a query. Views are one such facility that satisfies this requirement. Views allow a query result to be accessed as a virtual table, which simplifies the usage of complex queries. Figure 5 contains the view definitions for the Advertiser and Invoice views from the Advert schema. In this section we focus on materialized views and updatable views.

**Materialized Views:** A materialized view stores the results of a query on the disk as an object in the database, as opposed to being virtual and not stored. Another major difference is that a materialized view is not necessarily kept up-to-date with the data that it originates from. MSSQL supports materialized views indirectly, while Postgres uses a direct syntax. In order to implement what is similar to a materialized view in MSSQL, a unique clustered index must first be created on a view. These are referred to as Indexed Views [12]. In Postgres, a materialized view is created with a CREATE MATERIALIZED VIEW clause, which is then followed by the appropriate subquery [51]. Along with being created in a different manner, the two differ in that the MSSQL implementation keeps the information up-to-date with the original data. If the information should not be kept up to date, then it is always possible to create a new table in the database using the CREATE TABLE AS syntax. In Postgres, materialized views need to be manually updated with the REFRESH MATERIALIZED VIEW statement whenever necessary.

```
--Advertiser View
CREATE VIEW advertiser AS
SELECT a.advertiser_id, a.university_org,
a.nonprofit_org,
(SELECT SUM(ar.total_price)
 FROM ad_price ar
 WHERE ar.advertiser_id = a.advertiser_id)
 -
(SELECT SUM(p.amount)
 FROM payment p
 WHERE  p.advertiser_id  =  a.advertiser_id)
outstanding_balance
FROM advertiser_t a;

--Invoice View
CREATE VIEW INVOICE AS
SELECT
i.ad_id, i.invoice_date, ap.total_price,
CASE WHEN COALESCE(ap.total_price -
 (SELECT SUM(p.amount)
 FROM Payment p
 WHERE p.ad_id = i.ad_id),
 ap.total_price) > 0
 THEN 0 ELSE 1 END fully_paid,
CASE WHEN
 a.start_date + interval '1day'
 * a.num_issues < current_date THEN 1 ELSE
 0 END run_complete,
CASE WHEN a.start_date + interval '1day' *
 a.num_issues > current_date OR
 (SELECT SUM(COALESCE(ar.proof_sent, 0))
 FROM ad_run ar
 WHERE ar.ad_id = i.ad_id) = a.num_issues
 THEN 0 ELSE 1 END run_failed
FROM invoice_t i
JOIN ad a ON a.ad_id = i.ad_id
JOIN ad_price ap ON ap.ad_id = i.ad_id;
```
**Figure 5. View definitions of Advertiser and Invoice from the Advert schema**

**Updateable Views:** An updatable view allows write operations (INSERT, UPDATE, and DELETE) to be passed to the underlying tables in the view definition. Both MSSQL and Postgres support updateable views, albeit each with its own limitations. Views that `SELECT` data from one source table and have no computed columns are updatable in both DBMSs [15][62]. For example, the view `Advertiser` in the Advert schema is updateable because its definition selects data from only the `Advertiser_T` table. The query in Figure 6a inserts a new row into the `Advertiser_T` table, then updates and deletes the row through the `Advertiser` view.

Views defined using a `JOIN` cannot always be updated. MSSQL can perform `UPDATE` statements on these views in some situations, while Postgres will fail with an error message. Neither implementation permits row deletion on these views.

```
INSERT INTO ADVERTISER_T VALUES(2,0,0);

UPDATE ADVERTISER
SET advertiser_id = 3
WHERE advertiser_id = 2;

DELETE FROM ADVERTISER
WHERE advertiser_id = 3;
```
**Figure 6a. Example of using an updateable view (Advert)**

```
UPDATE INVOICE
SET invoice_date = '2017-02-01';

DELETE FROM INVOICE
WHERE ad_id = 0;
```
**Figure 6b. Example of a query attempting to update the `INVOICE` view (Advert)**

Figure 6b demonstrates performing `UPDATE` and `DELETE` statements on the view `INVOICE`, which is defined using two `JOIN` statements. The `UPDATE` in figure 6b will succeed in MSSQL, but fail in Postgres. The `DELETE` will fail on both DBMSs.

The documentation for both DBMSs goes into additional detail on the limitations of updatable views. The limitations in MSSQL are based on how the column being updated is derived. Any `INSERT`, `UPDATE`, or `DELETE` statement on a view can only modify columns that are derived from a single base table. Additionally, the column definition cannot include any aggregate functions [15]. For example, MSSQL is unable to update the `fully_paid` column in the `INVOICE` view (Figure 7).

```
UPDATE INVOICE
SET fully_paid = 0;
```
**Figure 7. Example query that updates a column in INVOICE derived from multiple tables (Advert)**

By comparison, Postgres' updatable views are more limited. The same limitations on updateable views in MSSQL apply to Postgres, however, Postgres makes the entire view not updateable, instead of just the offending column [62].

## 2.5 Querying Metadata

A necessary feature when creating or maintaining database designs is the ability to query metadata. Metadata stored by DBMSs includes many critical details about database objects. For example, metadata about a table includes the data type of each column. The ability to perform queries on metadata provides a method to evaluate the current implementation of a database. Metadata can be compared to design documents, and any discrepancies in the design and implementation can be addressed.

Both MSSQL and Postgres provide metadata access via proprietary "catalog views": MSSQL provides a total of 407 views in a schema named `sys` contained in the `master` database; Postgres defines 55 views in a per-database schema named `pg_catalog`. All MSSQL views span databases, but most Postgres views are database-specific, and in general, MSSQL provides more extensive metadata than Postgres. MSSQL contains the proprietary catalog view `sys.foreign_keys` and `sys.partitions`, for which Postgres has no direct substitute (although similar metadata can be retrieved using more complex queries). Even when the two systems provide comparable catalog views, the access mechanisms and the result schemas are often quite different. For example, Figures 8a and 8b show the queries to list all tables in the Advert schema. As the queries show, MSSQL requires a join not required in Postgres. An important detail not apparent in the queries is that the MSSQL result contains 31 columns whereas the Postgres result contains only 8 columns.

```
SELECT sys.tables.*
FROM sys.tables t JOIN sys.schemas s ON
t.schema_id = s.schema_id
WHERE s.name = 'Advert';
```
**Figure 8a. Catalog view query in MSSQL**

```
SELECT *
FROM pg_catalog.pg_tables
WHERE schemaname = 'Advert';
```
**Figure 8b. Catalog view queries in Postgres**

To overcome such differences among DBMSs, the SQL standard defines a metadata catalog called INFORMATION_SCHEMA (which we abbreviate to *info schema* for ease of writing). This catalog defines a total of 65 views with metadata for various database objects, including table columns. Figures 9a and 9b show two info schema queries—one to retrieve metadata for columns and another to retrieve metadata for tables—and each query works exactly as shown and returns the same result schema in both MSSQL and Postgres.

```
SELECT column_name, data_type,
  character_maximum_length
FROM INFORMATION_SCHEMA.COLUMNS
WHERE table_name = 'advertiser_t';
```
**Figure 9a. An `INFORMATION_SCHEMA` query (Advert)**

```
SELECT t.TABLE_NAME, c.COLUMN_NAME,
  c.CONSTRAINT_NAME
FROM
  INFORMATION_SCHEMA.TABLE_CONSTRAINTS t
JOIN
INFORMATION_SCHEMA.CONSTRAINT_COLUMN_USAGE c
ON t.CONSTRAINT_NAME = c.CONSTRAINT_NAME
WHERE CONSTRAINT_TYPE = 'FOREIGN KEY';
```
**Figure 9b. A more complex `INFORMATION_SCHEMA` query**

Both MSSQL and Postgres support the info schema, but Postgres' support is more complete. MSSQL only implements 21 info schema views, whereas Postgres implements 60 [36][58]. Additionally, Microsoft's documentation warns against using info schema to reliably determine the schema of a table [10]. As documentation for both DBMSs notes, the catalog views must be used instead of info schema to access DBMS specific metadata.

MSSQL's partial support for the info schema is a detriment to its standards compliance and makes writing portable code more difficult. While it does implement some of the more commonly used views, it is often more practical to use the catalog views. Thus, MSSQL does not support a portable method to query metadata. Postgres has more complete support for the info schema, making it more practical to use info schema over the catalog views.

## 3. DEVELOPMENT ACTIVITIES

Activities we categorized as *development* concern the logical and physical levels of schema management. These activities frequently involve the use of DML queries, and are heavily dependent on the SQL implementation of the DBMS. MSSQL and Postgres have many differences that require modified SQL code for the same effect in each implementation.

### 3.1 Common Table Expressions

Both MSSQL and Postgres support common table expressions (CTEs) using the `WITH` clause. For non-recursive expressions, the two systems are functionally identical. The recursive CTE syntax differs slightly. The SQL standard requires recursive CTEs to be explicitly declared using the `RECURSIVE` keyword [8§9.13]. MSSQL does not require the `RECURSIVE` keyword, however Postgres will only evaluate a CTE recursively if `RECURSIVE` is included. Figure 10 demonstrates a recursive `WITH` statement which generates a range of dates. This query works as-is in Postgres, and the MSSQL version can be obtained by removing the `RECURSIVE` keyword.

```
WITH RECURSIVE ISSUE_DATE_LIST AS (
  SELECT ad_id, start_date,
  CAST(num_issues AS int) ni
  FROM ad
  WHERE ad_id = (
    SELECT MAX(ad_id)
    FROM ad
  )
  UNION ALL
  SELECT ad_id, start_date, ni - 1
  FROM ISSUE_DATE_LIST
  WHERE ni > 1
)
SELECT ad_id, (start_date + ni)
issue_date
FROM ISSUE_DATE_LIST;
```

**Figure 10. Recursive CTE example (Postgres). MSSQL version is obtained by removing RECURSIVE on line 1.**

### 3.2 Query Batches and Transaction Boundaries

MSSQL and Postgres execute groups of queries that have been sent through a client interface differently. MSSQL executes groups of incoming queries in batches. The MSSQL documentation defines a query batch as, "a group of two or more SQL statements or a single SQL statement that has the same effect as a group of two or more SQL statements," [41]. When working in an interface such as SQL Server Management Studio, the DBMS interprets any script that results in the execution of two or more SQL statements as a query batch. MSSQL imposes some limitations on what groupings of statements can be in a single batch. Specifically, the following statements must be executed in their own batch: `CREATE DEFAULT`, `CREATE FUNCTION`, `CREATE PROCEDURE`, `CREATE RULE`, `CREATE SCHEMA`, `CREATE TRIGGER`, `CREATE VIEW`. Additionally, columns added or modified using an `ALTER TABLE` statement may not be referenced by other queries in the same batch [41].

MSSQL does not provide a native statement for explicitly separating multiple SQL statements into multiple batches. Instead, the client program must separate multiple statements into batches and separately send each batch the DBMS. To separate query batches, Microsoft's client programs implement the `GO` command [23]. When the program encounters a `GO` command, it sends all preceding commands in their own batch, and begins sending all following queries in a new batch. `GO` allows a single SQL script to execute multiple statements that are not allowed in the same batch.

Postgres does not require clients to separate queries into batches, however it does place some similar limitations on groupings on statements that can be executed in a single transaction. For example, `ALTER SYSTEM` and `CREATE DATABASE` statements must be executed in their own transactions.

### 3.3 Text Matching

Despite being a common task, the SQL standard defines limited support for text matching, in the form of the `LIKE` operator. This operator only supports two simple wildcard characters and the ability to say if the matching should be case sensitive. Both MSSQL and Postgres fully support the two wildcards, but MSSQL's `LIKE` operator performs only case insensitive matches. However, Postgres' performs case sensitive matches and provides the `ILIKE` operator for case insensitive matches.

Both MSSQL and Postgres provide additional functionality for text matching. In MSSQL, these matching patterns still use the `LIKE` keyword and allow the matching of a single character to a given set of characters [42]. Postgres' extension instead uses the `SIMILAR TO` and `NOT SIMILAR TO` keywords, which allow for additional metacharacters, such as checking for the repetition of a character a particular number of times, or matching either of two given expressions. Finally, Postgres also provides full support of POSIX regular expressions with the ~ operator, which can be negated or made case-insensitive with the addition of ! and * characters, respectively [56].

### 3.4 Performance Optimization

Both DBMSs use a cost-based system for determining the best execution plan for a query [24] [48]. The DBMS assigns a cost to each operation it can perform, and selects a set of operations that perform the requested query with the lowest cost. The documentation for both states that the DBMS generally selects the best plan, and advise against manually making changes to the plan [24] [48]. Despite this advice, both DBMSs provide facilities for optimizing query performance.

MSSQL provides *query hints* and *table hints*, which allow the developer to modify the behavior of the query optimizer. Query hints are added in an additional clause, and are only applied to that query. Each hint specifies how the query optimizer should perform a specific operation. Table 2 presents a partial list of query hints and a short description of each.

| Hint | Description |
| --- | --- |
| JOIN | Forces one or more specified join methods be used |
| UNION | Forces one or more specified union methods be used |
| GROUP | Forces one or more specified group by methods be used |
| FORCE ORDER | Forces joins to be performed in the order they appear in a query text |
| OPTIMIZE FOR | Forces the optimizer to generate a plan using a supplied value for one or more parameters instead of the actual value |
| MAXDOP | Forces the maximum degree of parallelism to the specified number |
| MAXRECURSION | Forces the maximum recursion depth to the specified number |

**Table 2. A partial list of MSSQL query hints**

Query hints also include a subset called table hints, which are applied to a single table in a query. This type of hint can force a specific index be used when accessing that table, or modify how the query plan will access the table.

Figure 11 contains an example of using the FORCE ORDER query hint. Using FORCE ORDER increases the cost of this query, because the query optimizer calculates that joining INVOICE before AD is more optimal.

```
SELECT ar.ad_id, ar.issue_date,
   a.premium_place, a.color, a.edit
FROM AD_RUN ar
JOIN AD a ON ar.ad_id = a.ad_id
JOIN INVOICE i ON ar.ad_id = i.ad_id
WHERE i.fully_paid = 1
OPTION(FORCE ORDER);
```

**Figure 11. Using a query hint (Advert)**

MSSQL provides an additional tool for optimizing performance called *plan guides*. Plan guides force the query optimizer to behave a certain way when it encounters a specific query text [28]. Plan guides are useful in a situation where the exact text of a query cannot be changed, such as when an external application is issuing the query. Plan guides are created using the stored procedure sp_create_plan_guide, which is provided the query text to match and the query hints that should be applied to it. Guides can only be scoped to a single database, so only plans from the current database a query is executing from are eligible to be applied to the query [28].

MSSQL provides an additional form of plan guides called *template plan guides*, which can apply plan guides even if the query text does not match exactly [28]. Template plan guides use a parameterized query, which allows them to match instances of a query with different parameter values. Template plan guides are created by first using sp_get_query_tempalte to get the query template for the desired query, and the using the procedure sp_create_plan_guide with the parameterized template returned by sp_get_query_tempalte.

By comparison, Postgres offers very few options for query optimization. Postgres does not provide a facility for changing the behavior of the optimizer on a query by query basis. Instead, Postgres exposes parameters that change the overall behavior of the optimizer [48]. Two types of configuration options are exposed, *method configurations* and *cost constants*.

Method configurations allow the user to disable certain query execution operations, similar to MSSQL's query hints. Unlike query hints, method configurations apply to all plans generated by the query planner (and thus all queries). The Postgres documentation warns that method configurations are not intended to be permanent solutions, rather they should only be used for performance testing [48]. Figure 12 shows how to globally disable hash joins using method configurations.

```
ALTER SYSTEM SET enable_hashjoin TO 'off';
SELECT pg_reload_conf();
```

**Figure 12. Example of Postgres method configurations**

Cost constants are the cost values of individual operations the query optimizer uses when selecting a plan. Postgres allows these constants to be changed to better suit the DBMSs query workload. For example, two cost constants are seq_page_cost and random_page_cost, which are the costs for sequentially and randomly fetching new pages, respectively. By default, sequential fetches have a cost of 1.0, and random fetches have a cost of 4.0. Changing these constants affects how the query optimizer chooses to read pages from disk. Lowering the value of random_page_cost relative to seq_page_cost would make the query processor less strongly prefer sequential access, while raising random_page_cost would cause the query processor to more strongly prefer sequential access.

### 3.5 Dynamic SQL

Both DBMSs support executing dynamically constructed strings as SQL queries. MSSQL provides the EXEC statement and sp_executesql stored procedure to accomplish this. EXEC compiles and executes a statement string at runtime, while sp_executesql takes a parameterized query string and the relevant parameters, constructs the final query, and then executes it.

Postgres also supports executing dynamic SQL using the EXECUTE statement. In addition, the format() function can be used to construct a parameterized query similarly to sp_executesql. The EXECUTE statement can only be used to execute dynamic SQL in a PL/pgSQL function. If the query is expected to return a single value, the syntax EXECUTE <query> INTO <var>; can be used to directly place the result of the query into a variable.

Figures 13a and 13b demonstrate the use of dynamic SQL in MSSQL and Postgres, respectively. Figure 13a shows how the EXEC statement can execute a query string, in this case an info schema query. Figure 13b shows how a statement that is normally not parametrizable, CREATE USER, can be parameterized using dynamic SQL.

### 3.6 Functions and Procedures

Both DBMSs provide functions and procedures. The term *stored procedure* is often used in literature, however we use the terms function and procedure for ease of writing. MSSQL supports separate procedures and functions, while Postgres implements procedures as a subset of functions [37] [65].

```
EXEC('SELECT table_name FROM
  INFORMATION_SCHEMA.TABLES');
```

**Figure 13a. Simple example of dynamic SQL in MSSQL (Advert)**

```
CREATE OR REPLACE FUNCTION
createUser(userName VARCHAR(63))
RETURNS VOID AS
$$
BEGIN
  EXECUTE format('CREATE USER %s
  ENCRYPTED PASSWORD %L', $1, $1);
END
$$
LANGUAGE plpgsql;
```

**Figure 13b. Parameterized dynamic SQL in Postgres (Advert)**

In MSSQL, functions are intended to perform an atomic set of operations on an input to provide some output, while procedures may execute several statements in one or more transactions. Procedures support input and output parameters, while functions support input parameters and return values. Procedures may call other procedures, however, functions may not call procedures, because a function must be an atomic operation [37]. Additionally, functions and procedures are created using separate statements (`CREATE FUNCTION` and `CREATE PROCEDURE`).

Postgres implements procedures as a special case of functions. Procedures are simply functions that may not return a value. Both are created using the `CREATE FUNCTION` statement. While both may contain many statements, functions and procedures may only contain a single transaction.

**Language Support**: SQL Server 2016 only allows T-SQL statements, or Microsoft .NET Framework common runtime language methods to be run in stored procedures [35]. MSSQL also provides support for running scripts written in the R programming language through SQL Server Machine Learning Services [33]. Additionally, the most recent version of MSSQL (SQL Server 2017) adds support for the Python programming language through this same framework [68] [29]. In contrast, Postgres allows user-defined function to be in of any language that has been implemented to run in Postgres. Postgres provides native support for C, and provides a framework to extend this to other languages, such as Python or R [59].

### 3.7 Exceptions in Functions and Procedures

Beyond creating functions and procedures, it is often necessary to respond to exceptional situations during their execution. MSSQL's exception handling follows the `TRY...CATCH` and `THROW` convention familiar to most computer programmers. If any execution error with a reported "severity" greater than 10 occurs during the processing of statements in the `TRY` block, control is passed to the `CATCH` block [38]. Although Postgres does not implement these `TRY...CATCH` blocks, the `EXCEPTION` keyword provides similar functionality. The `EXCEPTION` keyword can be used once in each `BEGIN` block (typically towards the end of the block). If an error occurs during the execution of a stored procedure, execution is moved to the statements following the `EXCEPTION` keyword. Here, conditions can be checked, followed by statements that can appropriately handle the exceptional circumstances [52].

### 3.8 Miscellaneous Activities

We also identified several minor activities related to development. While these activities do not individually represent the most important differences, they should be kept in mind during a schema migration process.

**Date Arithmetic:** MSSQL and Postgres provide different facilities for date arithmetic. MSSQL provides non-standard functions for performing date arithmetic, including `DATEADD()` and `DATEDIFF()` [18]. These functions take a date interval, such as day, and two `DATETIME` values and return the sum or difference of the two dates (Figure 14a). Postgres follows the standard, and allows for date arithmetic using normal arithmetic operators. Figure 14b performs the same query as Figure 14a in Postgres [57]. Note that DATEDIFF subtracts the first input value from the second, whereas Postgres arithmetic operator subtracts the second operand from the first.

```
SELECT DATEDIFF(day,
  (SELECT MIN(issue_date) first_issue
   FROM AD_RUN
   WHERE ad_id = 0),
  (SELECT MAX(issue_date) last_issue
   FROM AD_RUN
   WHERE ad_id = 0)) + 1 run_length;
```

**Figure 14a. Example of `DATEDIFF()` (Advert)**

```
SELECT
  (SELECT MAX(issue_date) first_issue
   FROM AD_RUN
   WHERE ad_id = 0) -
  (SELECT MIN(issue_date) last_issue
   FROM AD_RUN
   WHERE ad_id = 0) + 1 run_length;
```

**Figure 14b. Example of Postgres date arithmetic syntax (Advert)**

**Expressions in `ORDER BY` and `GROUP BY`:** Both MSSQL and Postgres allow an identical set of expressions to be a part of an `ORDER BY` or `GROUP BY` clause. If an expression uses a function, it must be a scalar function that performs an identical action to every value that is queried. As a part of the Babysitting schema, we were asked to query a list of birthdates and addresses, sorted by day of the year, in order to create a birthday mailing list. In order to perform the sorting, we used an expression in the `ORDER BY` clause that would sort the list by only the month and day parts of a date. Both the MSSQL and Postgres implementations used a single function to extract these date parts. However, they used different functions due to differences in the functions available to manipulate dates.

**`TRIM` Functions:** Postgres provides a `TRIM()` function which removes whitespace from both sides of a string. MSSQL does not implement a function called `TRIM()`. Instead, it implements `LTRIM()` and `RTRIM()`, which remove whitespace from the left and right sides of a string, respectively. When used together, they are effectively the same as `TRIM()`.

**MSSQL Stored Procedure Library:** MSSQL provides several built-in stored procedures for retrieving metadata from the system catalog. In particular, we discovered `sp_columns` when looking for a replacement for Oracle's `DESCRIBE` statement. `sp_columns` returns metadata about a table's columns, similar to the info schema query in figure 9a. Since MSSQL does not fully support info schema, `sp_columns` allows easy access to column metadata stored in the system catalog. Several other stored procedures with similar functionality are provided, such as `sp_tables` [31].

**`SELECT` Without `FROM`:** Another development activity we identified was using a `SELECT` statement to select data not inside the database. This feature is commonly used as a way to output text from a query without necessarily accessing data inside the database. Both MSSQL and Postgres accomplish this using `SELECT` statements without a `FROM` clause. This allows a query to output data that is not retrieved from the DBMS, such as a string literal in a SQL script.

**Type Casting and Conversion:** Since SQL is a strongly typed language, both MSSQL and Postgres provide facilities to convert between variable types. MSSQL provides the `CAST` and `CONVERT` statements for explicitly converting between data types. `CAST` provides the SQL standard facility for type conversion, using the syntax `CAST(<var> AS <type>)`. MSSQL will try to convert the supplied value to the requested data type, and throw an exception if the types are incompatible. `CONVERT` behaves similarly, and provides addition facilities for how the cast should be performed. For example, `CONVERT` can cast a `FLOAT` number into a `VARCHAR` with commas and a fixed fractional precision. MSSQL also provides `TRY_CAST` and `TRY_CONVERT`, which return `NULL` if the cast cannot be performed, instead of an exception. Additionally, the documentation provides a table of which data types can be implicitly converted, explicitly converted, or are incompatible [16].

Postgres also provides the `CAST` statement for type conversion, with the same syntax. Postgres will also attempt to implicitly convert types when necessary. It will first find if the operator or function has a version that matches the input type or types. If not, it will attempt to convert one or more of the input values to a compatible type. If one or more of the input values still does not match the required values, an exception is thrown. The same process is followed when attempting to insert values into a table [43].

**Grouping Sets:** The final minor development activity we found was advanced grouping through the use of `GROUPING SETS`. These include the two shorthands for common grouping sets: `CUBE` and `ROLLUP`. Grouping by a grouping set runs the `SELECT` expression on each column or expression in the grouping set, outputting the results for each element of the grouping set. This is an important feature because it allows this grouping to be performed with a single query, rather than relying on several `UNION ALL`s. Despite not being defined in the SQL standard, these are all implemented similarly in both MSSQL [16] and Postgres [53].

## 4. ADMINISTRATION ACTIVITIES

Activities we categorized as *administration* are those which involve the operations and maintenance of the DBMS, rather than those which deal with data modeling or querying. As we were only dealing with a single user and did not need to perform any maintenance on the systems, our experience with administration activities is limited to those involved during the setup of both DBMSs on our machines. However, we chose to also investigate other administration activities, including those performed via the Data Control Language (DCL) component of SQL, and other management and optimization activities.

### 4.1 Environment Setup

Microsoft provides a single Windows® installer for MSSQL [40]. This installer includes the sqlcmd [34] command-line utility and SQL Server Management Studio (SSMS), Microsoft's integrated graphical environment for managing instances of MSSQL.

In contrast, Postgres' website [67] does not directly provide a Windows installer and instead links to two installers provided by third-parties. One is an "Interactive Installer" provided by EnterpriseDB® [5], and the other is a "Graphical Installer" provided by BigSQL [1]. We found that there were relatively few differences between these two distributions. Finally, because Postgres is open source [66], there is also the option to create an individual or custom build of the software. Both installers came packaged with the psql [63] command-line utility. Although Postgres is not officially associated with a particular graphical interface, both distributions also included pgAdmin [69], a separate free and open source community project that provides a GUI for managing instances of Postgres.

### 4.2 Users and Roles

Although the SQL standard gives an outline of how a security and permissions system should be implemented, it leaves the separation between roles (groups) and users (authorization identifiers) to implementations. MSSQL provides a clear separation between groups and users, whereas Postgres merges them both into the same model (roles). However, they both offer a complete role-based access control system.

In MSSQL, the concept of roles always refers to groups. These roles can then be assigned to users. In terms of authorization identifiers, MSSQL requires the creation of a server-level login [30]. Following that, logins can then have database users assigned to them (one login per user per database) [17]. MSSQL also provides integration with Microsoft Windows accounts, allowing an easier infrastructure setup for organizations who already use a Windows authentication system [9].

In Postgres, a role may be either a user or a group. This is because roles can inherit from other roles. The `LOGIN` attribute is used to allow a role to establish a connection to the DBMS. The presence or lack of the `LOGIN` attribute can be seen as the equivalent of the separation between users and groups. For simplicity and backwards compatibility with scripts designed for earlier versions, Postgres also allows `CREATE USER` to be used instead of `CREATE ROLE`. Using `CREATE USER` automatically adds the `LOGIN` attribute to the role being created [50].

Figures 15a and 15b show example SQL statements to create roles and users, and associate users with roles. In the example, roles are created for Student and Instructor users. Following that, a student user is created, and a schema is also created for them. Appropriate permissions are then set on the schema for both the student and any user with the Instructor role. Note that the effective privileges that the student user has on the schema created for them differ between the two implementations. These simplified examples are based on our work in the ClassDB application [4]. More information on permissions can be found in the subsequent section.

### 4.3 Execution Permissions

Although a user with administrative or superuser rights is able to perform most management tasks without needing to switch roles, it is sometimes necessary to perform management tasks under a different role. More often, different permissions or restrictions are needed when executing a procedure, or when a procedure needs to impersonate another role in the database. Another reason to switch execution permissions would be as a part of following the least privilege principle, so users and applications will run with the minimum permissions necessary to complete a task, and only switch to a role with greater permissions when needed.

Both MSSQL and Postgres provide methods of doing changing execution permissions. In MSSQL, the EXECUTE AS <user> statement allows the execution context of a session to be changed [14], and Postgres permits changing the session and user identifier with the SET SESSION AUTHORIZATION <role> statement [64]. Both implementations also provide methods of undoing the context change. In MSSQL, the REVERT statement sets the execution context back to the caller of the previous EXECUTE AS statement, whereas in Postgres, using DEFAULT instead of <role>, or using the RESET SESSION AUTHORIZATION statement sets the context back to the originally authenticated user name. Postgres also allows the ability to only change the current role, and not the current session, through the SET ROLE <role> statement, and its correspondent: RESET ROLE. Different privilege checks are run depending on whether a session or a role is being switched. Postgres' implementation is closer to the implementation that the SQL standard recommends [8§14.5]. However, both MSSQL and Postgres allow changing the context during a transaction, which is a restriction suggested by the SQL standard. The Postgres documentation states that it "does not make this restriction because there is no reason to" [64].

```
--create roles and users
CREATE ROLE Student;
CREATE ROLE Instructor;
CREATE LOGIN Ramsey033;
CREATE USER Ramsey033;

--associate a user with a role
ALTER ROLE Student ADD MEMBER Ramsey033;

--create a schema object for illustration
CREATE SCHEMA ramsey033;

--give a particular user CRUD access to schema
GRANT SELECT, INSERT, DELETE, UPDATE
 ON SCHEMA ramsey033 TO Ramsey033

--let any member of Instructor role read
schema
GRANT SELECT ON SCHEMA ramsey033 TO Instructor;
```
**Figure 15a. Creating a student user and schema in MSSQL**

```
--create roles and users
CREATE ROLE Student;
CREATE ROLE Instructor;
CREATE ROLE ramsey033 LOGIN;

--associate a user with a role
GRANT Student TO ramsey003;

--create a schema object for illustration
CREATE SCHEMA ramsey003;

--give a particular user CRUD access to schema
GRANT ALL PRIVILEGES ON SCHEMA ramsey003 TO
ramsey003;

--let any member of Instructor role read
schema
GRANT USAGE ON SCHEMA ramsey003 TO Instructor;
```
**Figure 15b. Creating a student user and schema in Postgres**

**Function Execution Permissions:** Both DBMSs also provide facilities to set the default execution context of functions. This is useful in cases such as user activity journaling, since the user performing a journaled operation will likely not have direct access to the user activity journal. MSSQL allows the EXECUTE AS clause in function, procedure, and trigger definitions [21]. The EXECUTE AS clause is similar to the EXECUTE AS statement, however it instead causes the entire procedure to be executed in the specified context. There are three types of contexts the clause can use: CALLER, SELF, or <user> [21]. CALLER executes the trigger with the same permissions as the executing user. SELF executes the trigger with the same permissions as the trigger owner. As before, providing a specific <user> causes the procedure to execute in the specified context.

Postgres provides similar functionality with the SECURITY clause [60]. The SECURITY clause can only be used in a function definition. Additionally, since Postgres procedures are functions, and Postgres triggers execute exactly one function, the SECURITY clause can set their execution contexts as well. Two contexts types can be specified in the clause, DEFINER and CALLER. DEFINER executes the function in the context of the user that owns the function, while CALLER executes in the context of the executer.

### 4.4 Journaling and Monitoring User Activity

Another important DBMS activity is the ability to monitor user activity. Many applications have strict requirements on logging how users interact with the DBMS. This monitoring can take place at different levels, for example, logging connections to the DBMS or logging INSERT statements made on a specific table.

Both MSSQL and Postgres provide triggers to journal user activity. Triggers are facility that allows a procedure to be executed when some event happens, such as an INSERT or DELETE.

MSSQL supports three different types of triggers, while Postgres supports only two. Both DBMSs support triggers which trigger on DML or DDL events. MSSQL supports an additional type of trigger which executes when a user establishes a connection with the DBMS.

**DML Triggers:** Both DBMSs support the execution of triggers when DML statements are performed. The MSSQL documentation refers to this type of trigger as a DML trigger, while the Postgres documentation refers to them as regular triggers. Both MSSQL and Postgres require a DML trigger be executed against a single table or view. This implies that DML triggers only exist in the scope of a single database.

The trigger implementations in both DBMSs only support a subset of the standard functionality. Notably, MSSQL does not support row-level triggers, however equivalent behavior can be obtained using the inserted and deleted virtual tables [14] [61]. Neither MSSQL nor Postgres support the REFERENCING clause for specifying an alias for new or old rows [8§11.3]. Postgres also limits trigger procedures to a single user-defined function, instead of multiple statements [61]. Figures 16a and 16b each present a DML trigger for journaling inserts on the AD table.

**DDL Triggers:** Both DBMSs also support triggers that execute when DDL statements are performed. MSSQL refers to these as DDL triggers, while Postgres refers to them as event triggers. MSSQL can execute DDL triggers in the scope of a single database, or the entire server. Event triggers can only be executed in the scope of a database in Postgres.

```
CREATE TRIGGER updateAdJournal
ON AD
AFTER INSERT AS
BEGIN
   DECLARE @ip NVARCHAR(32) =
   (SELECT client_net_address
   FROM sys.dm_exec_connections
   WHERE session_id = @@SPID);
   INSERT INTO
   new_ad_journal(insert_timestamp,
   user_name, ip_address, ad_id)
   SELECT CURRENT_TIMESTAMP,
   ORIGINAL_LOGIN(), @ip, ad_id
   FROM inserted;
END
```

**Figure 16a. DML trigger example in MSSQL (Advert)**

```
CREATE OR REPLACE FUNCTION
updateNewAdJournal()
RETURNS trigger AS
$$
BEGIN
   INSERT INTO
   new_ad_journal(insert_timestamp,
   user_name, ip_address, ad_id)
   SELECT statement_timestamp(),
   session_user::TEXT,
   inet_client_addr(), NEW.ad_id;
   RETURN NEW;
END;
$$
LANGUAGE plpgsql;

CREATE TRIGGER updateAdJournalTrigger
AFTER INSERT
ON ad_t
FOR EACH ROW
EXECUTE PROCEDURE updateNewAdJournal();
```

**Figure 16b. DML trigger example in Postgres (Advert)**

**Logon Triggers:** MSSQL also supports logon triggers. These triggers are similar to DDL triggers, but are only executed when a user logs on to the database, and can only be executed in the scope of the whole server. Logon triggers support the same permissions mentioned previously. Postgres does not provide logon triggers, however it does have the ability to log connections to the external sever logs [49]. Accessing the external server logs can be prohibitive, as it requires knowledge of where the logs are stored, how they are formatted, access to the underlying server filesystem, and a procedure to import individual log files. Additionally, Postgres connection logs can take a considerable amount of space on the system, which increases proportionally to the number of connections made to the DBMS [3].

## 5. DISCUSSION

In this section, we discuss the overall findings of this study. We present a discussion of the standards compliance of both DBMSs, a comparison of activities supported, and detail what conclusions we draw from this study.

### 5.1 Summary of Standards Compliance

MSSQL is clearly less standards compliant than Postgres. MSSQL shows major deviations, such as its lack of support for the complete INFORMATION_SCHEMA, and minor deviations, such as allowing recursive CTEs without explicit declaration. While Postgres is largely standards compliant, there are several areas where both DBMSs deviate. For example, both deviate in their implementation of the CLOB data type. MSSQL uses a different syntax for performing the same functionality, while Postgres uses a different data type name.

Despite being largely standards-compliant, we still identify many differences between both DBMSs. The SQL standard primarily dictates features a standards-compliant DBMS must support. It often does not place strict limitations on the behavior of such features. One example is Postgres' greatly expanded range for the NUMERIC data type, as the standard does not define what the ranges must be. Another example is that the SQL standard requires support for views, but it does not define any requirements for the behavior of updatable views [8§4.3].

The SQL standard also allows both DBMSs to implement extended features that it does not define. Postgres' extensible native language support is an example of one such feature. Many of MSSQL's deviations from the standard are used to provide an extended feature set. For example, included columns and query hints are not defined in the standard, however these features are clearly useful for many applications. The clear disadvantage to these features is that they are not standardized. Therefore, an application depending on them becomes locked to platforms that support the required non-standard features.

### 5.2 Comparing Activities Supported

One conclusion that can be drawn from these differences is that standards compliant SQL code is not necessarily portable or optimal. While many of the differences shown may individually be minor, their sum can make schema migration difficult. For example, a schema that heavily relies on computed columns or updatable views may be difficult to migrate to Postgres, despite being otherwise standards-compliant. Likewise, a schema that relies on INFROMATION_SCHEMA may be difficult to migrate to MSSQL, and may require the use of non-standard features.

To help gauge the overall level of support both DBMSs showed for the activities we studied, we assigned a score for each activity based on the level of support the DBMS had for that activity. A score of two indicates complete, and, if possible, standards compliant support for that activity. A score of one indicates partial or non-standard support for that activity. A score of zero indicates no support for that activity. The score for each activity, as well as the total score for each DBMS can be seen in table 3.

The highest possible score was 52 points, a score of two in each of the 26 activities. MSSQL scores slightly higher than Postgres, at 42 points compared to 41. These scores indicate that overall, both DBMSs had a high level of support for the activities we studied. MSSQL tends to have at least partial support for many features, as shown by four scores of one and only three scores of zero. Conversely, Postgres tends to have no support instead of partial support, as evidenced by five scores of zero and only a single score of one. These different score distributions point to one of the fundamental differences between the activities supported by both DBMSs. MSSQL is more likely to have support for a given activity, however this support comes at the cost of standards compliance. Postgres has support for fewer activities, however those it does support are very likely to be standards compliant.

| Activity | MSSQL | Postgres |
|---|---|---|
| **Design** | | |
| Standard Field Types | 2 | 2 |
| Querying Metadata | 1 | 2 |
| Native Computed Columns | 2 | 0 |
| Updatable Views | 2 | 2 |
| Materialized Views | 1 | 2 |
| DML Triggers | 2 | 2 |
| Included Columns in Indexes | 2 | 0 |
| Filtered Indexes | 2 | 2 |
| Expressions in Indexes | 1 | 2 |
| **Development** | | |
| Common Table Expressions | 2 | 2 |
| Unrestricted Query Batches | 0 | 2 |
| Extended Text Matching | 1 | 2 |
| Dynamic SQL | 2 | 2 |
| Query Hints | 2 | 0 |
| Grouping Sets | 2 | 2 |
| Expressions in ORDER BY and GROUP BY | 2 | 2 |
| Standard Type Casting | 2 | 2 |
| **Administration** | | |
| DDL Triggers | 2 | 2 |
| Server Level Triggers | 2 | 0 |
| Role-Based Access Control | 2 | 2 |
| Execution Permissions | 2 | 2 |
| Full Linux Support | 0 | 2 |
| Stored Procedures | 2 | 1 |
| User Defined Exceptions | 2 | 2 |
| Plan Guides | 2 | 0 |
| Native Language Extensibility | 0 | 2 |
| **Total** | **42** | **41** |

**Table 3. Summary of support for each identified activity**

Using this evaluation, we can determine how suitable Postgres is for enterprise applications. The activities Postgres does support function very well, however, many enterprise applications require features that Postgres does not support, such as logon triggers or included columns in indexes. Additionally, there are few places where Postgres is technically more advanced than MSSQL. Some of these limitations can be overcome with additional development work, such as approximating a logon trigger using Postgres' connection logging feature. We must also consider that because Postgres' license [66] results in it being free and open source software (FOSS), it is possible to extend the DBMS itself and replace missing features.

Considering all these points, we can say that Postgres is suitable for enterprise applications from a technical standpoint, although the exact suitability to a specific application will depend on the required features. Also, even if Postgres is missing necessary features, it may still be viable since its FOSS nature makes it possible to extend. However, there may be hidden costs to using Postgres if a large amount of development work is needed to overcome its technical limitations.

## 6. SUMMARY

In this paper, we have presented a comparison of many common database design, development, and administration facilities in MSSQL and Postgres. We compared what level of support, if any, both DBMSs had for each facility. We also compared the implementation of each facility to the SQL standard, if applicable. We then presented a quantitative measure of each DBMSs overall performance by assigning a score to the level of support and standards compliance for each facility.

Based on our analysis of 26 common database activities, we conclude the following: Both MSSQL and Postgres include competitive features to support common database activities involved in building enterprise systems, with MSSQL taking a more "implementer friendly" approach. Postgres is quite suitable (technically speaking) for enterprise applications, but its lack of certain features may mean some activities require additional implementation effort. Lastly, Postgres's SQL implementation is very close to the SQL standard, whereas MSSQL is not. However, standards-compliant code is often not portable and at times can be quite inefficient.

We feel Postgres is suitable for enterprise applications from a technical standpoint, although its exact suitability to a specific application will depend on the required features. Also, even if Postgres is missing necessary features, it may still be viable because its FOSS nature makes it possible to extend. However, there may be hidden costs to using Postgres if a large amount of development work is needed to overcome its technical limitations.

## ABOUT THE AUTHORS

Andrew Figueroa and Steven Rollo are undergraduate Computer Science students at the Western Connecticut State University (WCSU). Dr. Sean Murthy is a member of the Computer Science department at WCSU, and also heads the Data Science & Systems Lab (DASSL) at WCSU.